\documentclass[lettersize,journal]{IEEEtran}
\usepackage{amsmath,amsfonts}
\usepackage{algorithmic}
\usepackage{algorithm}
\usepackage{array}
\usepackage[caption=false,font=normalsize,labelfont=sf,textfont=sf]{subfig}
\usepackage{textcomp}
\usepackage{stfloats}
\usepackage{url}
\usepackage{verbatim}
\usepackage{graphicx}
\usepackage{cite}
\usepackage[table,xcdraw]{xcolor}
\colorlet{shadecolor}{yellow}
\usepackage{color,soul}

\usepackage[justification=centering]{caption}
\begin{document}

\title{Flexible Spectrum Orchestration of Carrier Aggregation for 5G-Advanced}

\author{Xianghui Han, Chunli Liang, Ruiqi Liu, Xingguang Wei, Mengzhu Chen, Yu-Ngok Ruyue Li, Shi Jin}

\maketitle

\begin{abstract}
With increasing availability of spectrum in the market due to new spectrum allocation and re-farming bands from previous cellular generation networks, a more flexible, efficient and green usage of the spectrum becomes an important topic in 5G-Advanced. In this article, we provide an overview on the 3rd Generation Partnership Project (3GPP) work on flexible spectrum orchestration for carrier aggregation (CA). The configuration settings, requirements and potential specification impacts are analyzed. Some involved Release 18 techniques, such as multi-cell scheduling, transmitter switching and network energy saving, are also presented. Evaluation results show that clear performance gain can be achieved by these techniques.

\end{abstract}


\section{Introduction}\label{sec1}

\IEEEPARstart{T}{he} rapid development of the fifth generation (5G) wireless communication systems have empowered a more digitized and more diverse society with supporting more stringent requirements in terms of bandwidth, date rate, latency and reliability etc. In the first three releases (i.e., Release 15$ \sim $17) of 3rd Generation Partnership Project (3GPP) 5G new radio (NR), many techniques have been exploited to meet some standalone requirements, e.g., downlink (DL) throughput, or to meet some specific compound requirements, e.g., ultra-reliable and low latency communication (URLLC) \cite{ref1}. Stepping into 5G-Advanced era from NR Release 18, it aims to achieve more comprehensive requirements raised by new emerging use cases such as machine vision and eXtended Reality (XR) and so on \cite{ref2}. A smarter and greener society will come along with the development of 5G-Advanced. 

 Carrier aggregation (CA) is one of the most powerful techniques that is used in wireless communication system to increase data rate per user equipment (UE), whereby multiple component carriers are assigned to the same UE \cite{ref3}. CA has been widely used in both Evolved Universal Terrestrial Radio Access (E-UTRA) and NR for support of wider bandwidth. One straightforward solution for CA is to aggregate continuous component carriers within the same operating frequency band, i.e., intra-band CA. However, this may not always be possible, due to the frequency allocations of one operator. For non-continuous CA, it could either be intra-band, i.e., the component carriers belong to the same band while with frequency gap(s) or it could be inter-band, in which case the component carriers belong to different operating frequency bands. CA can support both frequency division duplex (FDD) and time division duplex (TDD) bands. That is, FDD-FDD CA, TDD-TDD CA and  FDD-TDD CA are all supported in the current 3GPP specifications. In NR Release 15$ \sim $17, some enhanced CA operations have been specified, such as cross cell scheduling, uplink (UL) transmitter (Tx) switching, fast secondary cell (SCell) activation etc \cite{ref4}\cite{ref5}. CA enhancements also attract abundant study in academia on various aspects such as power control and scheduling optimization etc\cite{ref6}\cite{ref7}. 

However, the existing CA framework is still not flexible enough to cater for new scenarios or use cases. For instance, some operators may have several scattered spectrum resources with narrow bandwidth at 700/800/900 MHz, it is desirable to use these spectrum resource in a more spectral/power efficient manner by removing the transmission of DL synchronization signals if possible. Another example is, some operators may have a band with restrictions due to regional regulations, e.g., limited power or indoor usage for 2.3GHz band in China. The existing CA framework mainly has two restrictions to achieve such targets. The first restriction is the UL and DL carrier frequency for one serving cell has to be from the same band. The other is one UL component carrier is always associated with one DL component carrier, and one DL component carrier cannot be associated with multiple UL component carriers. In 3GPP RAN\#93 e-meeting, flexible spectrum access has been discussed, where one of the most promising ways is CA based framework with flexible association of DL and UL carriers \cite{ref8}. Another alternative is multi-band serving cell which is based on the extension of supplementary UL (SUL). Considering SUL is only targeted for better UL coverage while having limitations on improvement of UL capacity, e.g., due to not allowing simultaneous transmission of normal UL and SUL, this paper focuses on the enhancements via CA framework for support of more use cases and scenarios.

In addition, network energy saving is of great importance for environmental sustainability, to reduce environmental impact, and for operational cost savings \cite{ref9}. 5G can significantly improve the user experience with higher throughput, massive connections, and various services. However, the energy consumption caused by the 5G base stations and user devices have become an acute issue with the deployment of 5G. Most of the energy consumption comes from the radio access network \cite{ref10}. The power consumption of radio access can be split into two parts: the dynamic part which is only consumed when data transmission/reception is ongoing, and the static part which is consumed all the time to maintain the necessary operation of the radio access devices even when the data transmission/reception is not on-going. Therefore, limiting DL channels/signals transmission (e.g., in SCell) is a promising solution for power saving in CA scenarios from network perspective. This can reduce the dynamic power consumption part or the network can even choose to shut down the whole DL operation of an Scell if no any DL transmission on it. In Release 18 network energy saving work item, SCell without Synchronization Signal and Physical Broadcast Channel Block (SSB) for inter-band CA is proposed to save the network energy \cite{ref11}.

In this paper, spectrum orchestration with support of flexible association of DL and UL carriers under CA framework is discussed, with configuration settings, requirements and potential specification impacts demonstrated. Some key techniques that enable flexible associated CA in a more efficient and greener manner are analyzed, including one donwlink control information (DCI) scheduling multiple Physical Downlink Shared Channel (PDSCH)/Physical Uplink Shared Channel (PUSCH), UL Tx switching among up to 4 bands with 2 Tx, and SCell without SSB for inter-band. Evaluation results for these techniques are also provided and clear performance gain is observed. With thorough analysis, the capability of supporting new use cases and achieving network energy saving via flexible associated CA in 5G-Advanced can be then envisioned.

The remaining parts of this article are organized as follows. We first discuss the framework of flexible associated CA in Section \ref{sec2}. And in Section \ref{sec3}, some key techniques enabling flexible associated CA framework are presented. Section \ref{sec4} provides the evaluation results for these techniques. Finally, Section \ref{sec5} concludes this article.

\section{Framework of flexible association of DL and UL carriers}\label{sec2}

\begin{figure}[ht]
  \begin{center}
  \includegraphics[scale=0.5]{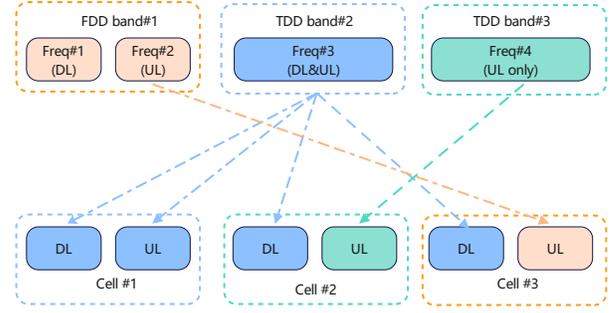}\\
  \caption{Enhanced CA framework with flexible association of DL and UL carriers}
  \label{fig_1}
  \end{center}
\end{figure}

To achieve flexible spectrum orchestration, an enhanced CA framework with flexible association of DL and UL carrier is depicted in Fig.\ref{fig_1}. It mainly includes the following new characteristics compared to the existing CA framework.
\begin{itemize}
   \item{The DL and UL carriers of one serving cell can come from different frequency bands.}
   \item{One DL carrier can be associated with different UL carriers from different serving cells, or vice versa.}
   \item{One serving cell may only have UL carrier(s), i.e., UL SCell only.}
\end{itemize}

Before going into the details, it is necessary to first provide clear definition for frequency band, frequency carrier and serving cell. A frequency band is a range of frequencies in a spectrum which is defined by 3GPP RAN4 in TS38.101\cite{ref12}.  Initially, frequency bands are categorized into FDD bands and TDD bands. But with the introduction of supplementary DL (SDL) and SUL, SDL bands and SUL bands are also defined. An FDD band contains at least a UL/DL carrier pair, while a TDD band contains at least a bidirectional carrier. An FDD cell includes an uplink carrier and a downlink carrier both from one FDD band, and a TDD cell includes a bidirectional carrier in one TDD band. In case of SUL configured, a serving cell (for both FDD cell and TDD cell) includes at most one SUL carrier and one non-SUL carrier. For a cell, it can be configured with a downlink carrier only, but it cannot be configured with an uplink carrier only in the current 3GPP specifications.

\begin{figure}[ht]
  \begin{center}
  \includegraphics[scale=0.55]{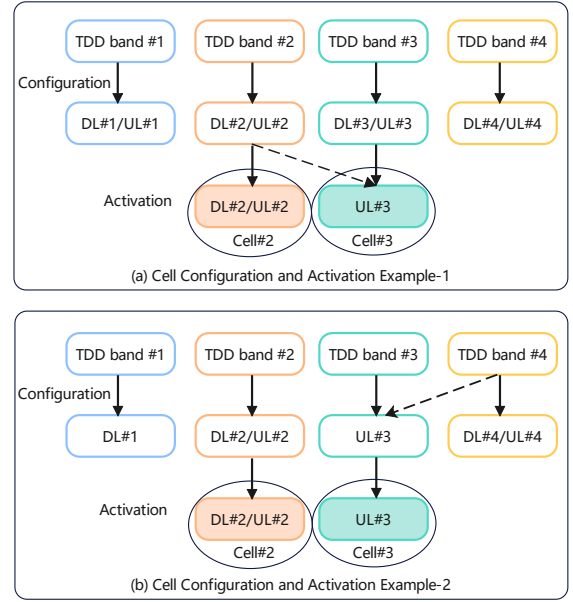}\\
  \caption{Two examples for cell configuration and activation for the enhanced CA framework}
  \label{fig_2}
  \end{center}
\end{figure}

For the enhanced CA framework, an example is shown in Fig.\ref{fig_1}, where three cells are generated by three different frequency bands respectively. Band \#1 is an FDD band, which includes a DL carrier (Freq \#1) and a UL carrier (Freq \#2). Band \#2 is a TDD band with a bidirectional carrier (Freq \#3). And Band \#3 is a TDD band with only UL transmission allowed due to regulations. The cell configurations break the limitations of the existing CA framework. As illustrated in Fig.\ref{fig_1}, the following three cells are configured.

\begin{itemize}
    \item Cell \#1 is configured with the bidirectional carrier in TDD band \#2. This is a traditional TDD cell.
    \item Cell \#2 is configured with a DL carrier in TDD band \#2 and a UL carrier in TDD band \#3. This is an enhanced cell which is generated from two bands.
    \item Cell \#3 is configured with a DL carrier in TDD band \#2 and a UL carrier in FDD band \#1. This is also an enhanced cell which is generated from two bands.
\end{itemize}

Based on this enhanced CA framework, more flexible association of UL and DL carriers is supported for satisfying different requirements. For example, if the UE is configured with Cell \#1 and Cell \#2 for CA operation, it can support more uplink carriers than downlink carrier, which is suitable for UL-heavy scenarios. If the UE is configured with Cell \#2, it is suitable to handle special carrier requirements in one band due to regulation or spectrum sharing, e.g., Band \#3 is a frequency band at 2.3GHz for spectrum sharing by restricting to UL only transmission in outdoor deployment so that interference with other co-existed systems can be controlled. If the UE is configured with Cell \#3, it can satisfy the requirements of both higher downlink throughput and larger uplink coverage, since the TDD band is typically a band with larger bandwidth to increase the data throughput while the FDD band is typically a band with lower frequency band for coverage improvement.

\subsection {Cell Configuration and Activation for Enhanced CA Framework}

The flexible association of UL and DL carriers can be realized by proper setting of the cell configuration and activation. The following settings can be considered as candidate solutions.

\begin{itemize}
    \item Setting \#1: Configure legacy cells and activate flexible associated CA cells.
    \item Setting \#2: Configure flexible associated CA cells and activate cells without changing the carrier association for any cells. 
\end{itemize}

For Setting 1 as shown in Fig.\ref{fig_2}(a), four legacy cells are configured within 4 different bands. During cell activation procedure, only Cell \#2 and Cell \#3 are activated. For Cell \#3, it could be an UL-only SCell as shown in the example or it could also be a cell with both DL carrier from band \#2 and UL carrier from band \#3. This allows the network to dynamically activate flexible cells, e.g., a cell with only DL carrier or only UL carrier from a cell configured with both DL and UL depending on the varying traffic in time. For Setting \#2 shown in Fig.\ref{fig_2}(b), during cell configuration procedure, Cell \#1 is configured as a DL only cell with DL carrier from TDD band \#1. Cell \#3 is an UL only cell with UL carrier from TDD band \#3 or alternatively associated with a DL carrier from TDD band \#4. Cell \#2 and Cell \#4 are configured as legacy cells. During cell activation procedure, only Cell \#2 and Cell \#3 are activated in the example. For Setting \#2, the activation procedure would not allow to change the configuration of a particular cell. The advantage of Setting 1 is to provide a more flexible activation mechanism. On the other hand, Setting 2 is beneficial in terms of lower configuration overhead. 

\subsection {DL-limited Cell or UL-only Cell}

In NR, a UE receives Primary Synchronization Signal (PSS) and Secondary Synchronization Signal (SSS) in order to perform cell search. After cell search procedure, the UE needs to receive Physical Broadcast Channel (PBCH) to obtain the necessary system information for the subsequent reception/transmission. The Synchronization Signal (SS) and PBCH are packed as a single block called SSB as a whole. The SSB is the basis for a UE to access the network. On the other hand, continuously transmitting SSB in all serving cells in CA scenario causes large signaling overhead and unnecessary energy consumption. In NR up to Release 17, SCell without SSB for intra-band CA is supported. In Release 18, it is under discussion on whether or not to support SCell without SSB for inter-band CA in network energy saving work item. 

\subsubsection {Requirements of SCell without SSB for inter-band CA}
As specified in NR RAN4 specification TS38.133, the following requirements are defined if SSB is not provided for the SCell being activated in case of intra-band contiguous CA \cite{ref13}.
\begin{itemize}
    \item The received time difference (RTD) between the target SCell and the contiguous active serving cell is within ±260 ns.
    \item The difference of the reception power with the contiguous active serving cell is no larger than 6 dB.
    \item The reference signal(s) of SCell being activated has quasi co-location (QCL)-TypeA of relationship with temporary reference signal(s) (TRS(s)) of the SCell being activated, and the TRS(s) of the SCell being activated has (have) further QCL-TypeC relationship with SSB(s) of any active serving cell that is contiguous to the SCell being activated on that frequency range 1 (FR1) band.
\end{itemize}

For inter-band CA in Release 18 discussion, it is more reasonable to first focus on co-site scenario only, in which case the UL only SCell is assumed to be co-site with the special cell (SpCell), which includes primary cell of master cell group (MCG) or secondary cell group (SCG), or other normal SCell and the co-site cells are assumed to be synchronized with each other. In such case, it is observed the residual timing alignment error among different bands can be much less than 260 ns which is mainly dominated by the group delay of power amplifier (PA) and filtering in different bands in case of shared RF transceiver or additionally residual transceiver alignment error in different bands in case of separated RF transceiver. For the second requirement on power difference, it can be easily satisfied if the carrier frequencies of two bands are not far from each other in co-site scenario. If there is still TRS transmission in the SCell for inter-band CA, the last requirement above can also be ensured as the situation is the same as that of intra-band CA case. 

\subsubsection {Potential specification impacts of UL-only SCell}

There are many uplink signals/channels transmitted on a cell and many uplink related procedures defined for a cell. If these signals or channels are transmitted on an UL-only cell, we should first identify whether they can be transmitted properly in a cell without any downlink transmission.

\begin{itemize}
    \item Sounding Reference Signal (SRS)/PUCCH transmission
    \begin{itemize}
        \item General SRS/PUCCH configuration only depends on the UL carrier.
        \item The pathloss estimation for power control for SRS/PUCCH depends on the SSB or CSI-RS, which can be from PCell.
        \item The source reference signal for SRS/PUCCH spatial relationship configuration can be from other cell.
    \end{itemize}
   
    \item Dynamic grant (DG)-PUSCH:
    \begin{itemize}
        \item General PUSCH configuration only depends on the UL carrier.
        \item The pathloss estimation for power control for PUSCH depends on the SSB or CSI-RS, which can be from PCell.
        \item For codebook based PUSCH and non-codebook based PUSCH transmission, the spatial relationship can be indicated via the SRS indicator in the scheduling DCI.
         \item PUSCH can be scheduled by cross-carrier scheduling.
    \end{itemize}    

    \item Configured grant (CG)-PUSCH:
    \begin{itemize}
        \item Type 1 CG PUSCH only depends on the RRC configuration for the UL carrier.
        \item Type2 CG PUSCH depends on RRC configuration and DCI activation, which can be realized by cross-carrier activation.
        \item The other parts are similar as DG PUSCH.
    \end{itemize}   
    
    \item Timing advance (TA) maintenance 
    \begin{itemize}
        \item There will always be a normal SCell/SpCell (i.e. with DL) in the same TA group (TAG) of UL only SCell, and the UL only SCell can only be set to active in case there is active SCell/SpCell in the same TAG. This can be ensured by network implementation.
        \item The TA will be maintained based on the normal SCell/SpCell in the same TAG.
    \end{itemize}
    
    \item Radio resource management (RRM) 
    \begin{itemize}
        \item In co-site scenario, the addition/change/release of UL only SCell will be performed based on the measurement on the co-site normal SpCell/SCell, and this can be left to network implementation.
        \item No measure object will be configured for the UL only frequency.
    \end{itemize}
    
    \item Cross-carrier scheduling 
    \begin{itemize}
        \item The existing cross-carrier scheduling framework can be reused. It can keep some of the \emph{SearchSpace} configurations in the UL-only cell as legacy procedure, or alternatively include the necessary configurations (e.g., \emph{SearchSpace} configurations) in the scheduling cell as a new procedure.
    \end{itemize}
    
   \item TDD configuration 
    \begin{itemize}
        \item The existing TDD configuration supports configuring all symbols/slots as UL.
    \end{itemize}
    
    \item SCell activation procedure
    \begin{itemize}
        \item The existing SCell activation depends on the SSB or TRS. The procedure requires updates to support activation for UL-only SCell. For instance, MAC-CE triggers UL-only SCell activation and potential SRS/PRACH transmission for potential TA adjustment and early UL channel measurement.
    \end{itemize}
    
\end{itemize}

Based on above analysis, it can be concluded that the configuration of UL only SCell can be supported in the current RRC signaling. New UE capability needs to be defined for the UL only SCell, and it needs to update current SCell activation procedure for UL-only SCell.

\section{Key techniques for flexible associated CA}\label{sec3}

In this section, some key techniques that can be applied to the flexible associated CA framework and being discussed in NR Release 18 are provided. 

\subsection{Multi-Cell (MC) Scheduling}\label{sec3-1}

The motivation of the introduction of MC scheduling is to save downlink control resources in order to alleviate Physical Downlink
Control Channel (PDCCH) congestion \cite{ref14}. This is particularly useful in case of UL only SCell or asymmetric CA with more UL carriers than DL carriers, in which scenarios the DL resources are limited while it requires more DL resources to schedule UL transmissions. 

As an example, one DCI can be used to schedule three PUSCHs in three different cells via MC scheduling instead of using three independent DCI. It is expected that performance gain could be obtained as long as the size of a joint DCI for MC scheduling is smaller than the total size of multiple single-cell(SC) DCIs. For MC scheduling, the main items that need to be discussed in Release 18 includes the followings. 

\begin{itemize}
    \item Determination of the maximum number of PUSCHs and PDSCHs across different cells can be scheduled by one MC-DCI.
    \item MC scheduling applicable scenarios, e.g., sub-carrier spacing (SCS) assumption across cells, supported carrier type, whether to support SCell scheduling PCell and whether to support both MC-DCI and legacy DCI for the same scheduled cell on the same or different scheduling cells. 
    \item MC-DCI monitoring enhancement, such as determination of DCI size budget for MC-DCI and blind decoding/monitoring control channel element size budget etc. 
    \item MC-DCI information field discussion to determine which fields should be shared or separated for different scheduled cells.
    \item Hybrid automatic repeat request acknowledgement (HARQ-ACK) enhancements for MC-DCI scheduled PDSCHs. 
\end{itemize}

\subsection{UL Tx Switching}\label{sec3-2}

UL Tx switching is an important technique introduced for multi-carrier operation. It aims to select the better UL carrier and better transmission mode in a given slot for UL transmission via up to two con-current Tx. In Release 16, it is assumed that only one Tx can be supported on the first frequency band and two Tx can be supported on the second band by Tx switching from the first band. With more bands supporting two Tx, two Tx transmission on each of the two bands is specified in Release 17 by switching of both Tx between two bands. However, there are some limitations of current mechanism, e.g. UE can only be configured with at most 2 UL bands, which can only be changed by radio resource control (RRC) (re)configuration, and UL Tx switching can be only performed between 2 UL bands for 2 Tx capable UE. Therefore, UL Tx switching across up to 3 or 4 bands with up to 2 Tx simultaneous transmission is under study in Release 18 \cite{ref14}. This technique is also very helpful for support of the enhanced CA framework as more uplink carriers from more than 2 UL bands could be the typical scenarios for UL heavy use cases via flexible carrier association.

In NR Rel-16/17, there are two options supported for UL Tx switching, namely, \emph{'switchedUL'} and \emph{'dualUL'}. In the option of \emph{'switchedUL'}, simultaneous transmissions on 2 bands are not supported, which results in UL Tx switching equivalent to band switching. And for option of \emph{'dualUL'}, simultaneous transmissions on 2 bands are supported.

To support UL Tx switching with up to 3 or 4 bands, there could be two potential frameworks to achieve this.

\begin{itemize}
    \item Framework \#1: Dynamic Tx carrier switching can be performed across all the carriers in all configured 3 or 4 bands by the UE based on the UL scheduling,
    \item Framework \#2: Network indicates 2 bands out of the configured 3 or 4 bands via DCI or MAC-CE, and dynamic Tx carrier switching is between indicated bands the same as Rel-17. 
\end{itemize}

\begin{figure}[ht]
  \begin{center}
  \includegraphics[scale=0.45]{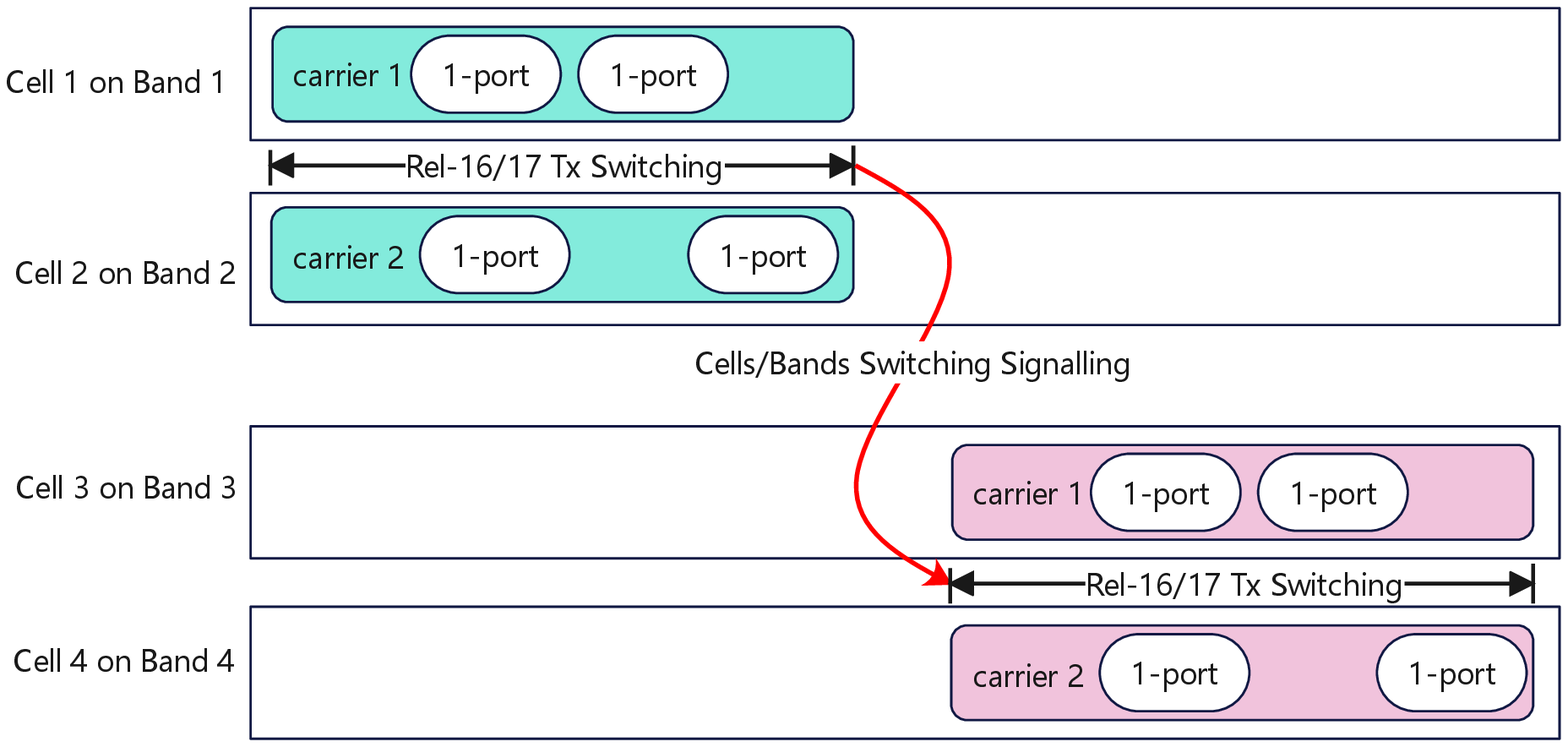}\\
  \caption{Example of Framework \#2, i.e., dynamic Tx carrier switching among two indicated bands}
  \label{fig_3}
  \end{center}
\end{figure}

Framework \#1 can be regarded as the extension of the Rel-16/17 UL Tx switching cases, where new tables for more Tx switching cases for 3 or 4 bands are first defined respectively and then identify each switching case and potential ambiguity issues respectively. Framework \#2 is designed with the intention of reusing Rel-16/17 UL Tx switching as much as possible. With framework \#2, gNB first indicates the cells within up to 2 bands for subsequent UL transmission, and then Rel-16/17 UL Tx switching can be performed within the 2 bands until next gNB indication. An example of Framework \#2 for Tx switching among 4 bands is shown in Fig.\ref{fig_3}.

Base on the Framework \#2, the legacy Rel-16/17 UL Tx switching can be reused including the following.
\begin{itemize}
    \item Legacy tables can be reused between the 2 indicated bands.
    \item The legacy switching cases can be reused.
    \item Legacy solution for ambiguity issues can be reused.
\end{itemize}

\subsection{Energy saving schemes for CA}\label{sec3-3}
For the connected-mode UEs configured with CA, the base station (BS) can perform the DL/UL transmission on each cell. For the SCell, when there is low DL traffic load or even no DL transmission required, SSB transmission takes up most of the DL transmission occasions, and the BS cannot enter into a sleep mode for energy saving \cite{ref9}. In such situation, the power consumption of SSB transmission results in a significant amount of network power consumption. It implies that limited or even no SSB transmission in SCells is an important solution to reduce base station power consumption, on top of the benefits of saving signaling overhead and using UL-only bands as discussed in Section \ref{sec2}.

For inter-band CA, the RF chains and other processing units between different cells are decoupled. Therefore, if the SSB transmission in the inter-band SCell can be reduced or even eliminated, the corresponding components can be muted so that network can obtain more energy saving benefits. In Section \ref{sec2}, the requirements and potential specification impacts of SSB-less SCell for inter-band have been presented. 

Considering the impact of network energy saving techniques on system performance (e.g., throughput, latency, etc.) and user experience, an energy saving fallback mechanism using an uplink wake-up mechanism can be promising. Specifically, when the network enters into an energy saving state, e.g., no any DL transmission, the time/frequency error of UE may be impacted. In this case, a wake-up signal (WUS)  transmitted from UE to network can be introduced to meet the flexible service requirements and minimize the impact on user experience. After receiving the WUS, the network can adjust the SSB transmission to respond to the UE’s indication. 

\section{Evaluation Results}\label{sec4}
\subsection{Evaluation for MC scheduling}\label{sec4-1}

In the simulation, the DCI size for single-cell scheduling is assumed as 60 bits excluding CRC. $N$ DCIs with each size of 60 bits are needed when the number of scheduled cells is $N$. For MC scheduling, it is assumed that the DCI size would be increased by 12 bits for each additionally scheduled cell. That is, to schedule $N$ cells in MC scheduling, a DCI with size of $ (60+12\times (N-1))$ is needed.

\begin{figure}[ht]
  \begin{center}
  \includegraphics[scale=0.6]{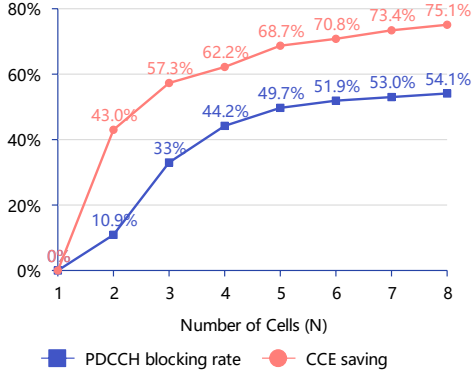}\\
  \caption{Gain of PDCCH blocking rate and CCE saving Vs Number of scheduled cells}
  \label{fig_4}
  \end{center}
\end{figure}

The gain of PDCCH blocking rate and control channel element (CCE) saving is presented in Fig.\ref{fig_4}. As shown, both the DCCH blocking rate gain and the CCE saving gain will increase as the number of scheduled cells increases. But the incremental gain becomes relatively small when the number of scheduled cells is larger than 5. On the other hand, the maximum size of DCI by the Polar code is 140 bits in current specification. Thus, it may be a good choice to set the maximum number of scheduled cells to 4 considering both the restriction of the DCI encoded by Polar code and the optimal performance of multi-cell scheduling. 

\subsection{Evaluation for UL Tx switching}\label{sec4-2}
In this subsection, the system level evaluations for UL Tx switching 4 bands for both Framework \#1 and Framework \#2 are presented. The band configurations for the 4 bands are shown below.
\begin{itemize}
    \item Band 1: a TDD band with carrier frequency 
    $@$4GHz, SCS = 30 kHz, BW = 100MHz, and the frame configuration with 'DDDSUDDSUU', and the configuration in special slot with '10 DL : 2 Gap : 2 UL' symbols.
    
    \item Band 2: a TDD band with carrier frequency 
    $@$2.6GHz, SCS = 30 kHz, BW = 100 MHz, and the frame configuration with 'DDDDDDDSUU', and the configuration in special slot with '6 DL : 4 Gap : 4 UL' symbols.
    
    \item Band 3: a FDD band with carrier frequency 
    $@$700MHz, SCS = 15 kHz, BW = 20 MHz.
    
    \item Band 4: a FDD band with carrier frequency 
    $@$2GHz, SCS = 15kHz, BW = 20 MHz.    
\end{itemize}

\begin{figure}[ht]
  \begin{center}
  \includegraphics[scale=0.6]{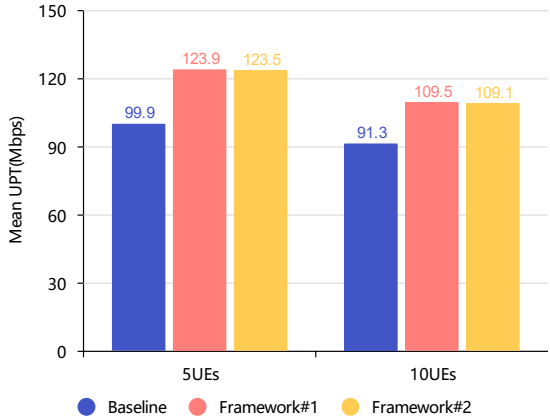}\\
  \caption{UPT Performance for UL Tx Switching with 4 bands}
  \label{fig_5}
  \end{center}
\end{figure}

The baseline is the performance of UL Tx switching with 2 bands. The mean user perceived throughput (UPT) is used as the evaluation metric as provided in Fig.\ref{fig_5} for 5 UEs and 10 UEs. As observed, in case of 5 UEs, the performance gain of Framework \#1 and Framework \#2 over the baseline in terms of mean UTP is about 23.93\% and 23.61\% respectively. And the performance loss of Framework \#2 compared to Framework \#1 is only about 0.28\%. Similar observation can be obtained in case of 10 UEs per cell. On the other hand, as discussed in Section \ref{sec3-2}, Framework \#2 has some advantages on specification efforts and forward compatibility. 

\subsection{Evaluation for network energy saving}\label{sec4-3}

\begin{figure}[ht]
  \begin{center}
  \includegraphics[scale=0.6]{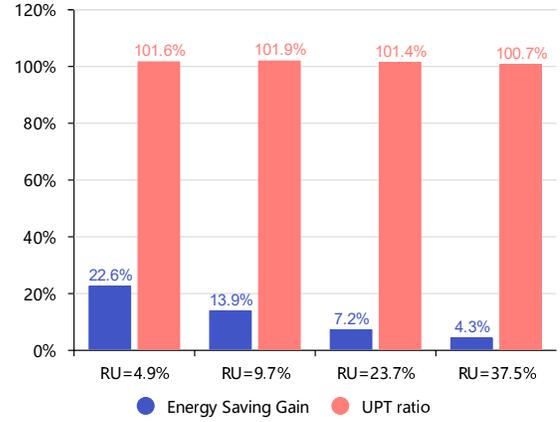}\\
  \caption{Evaluation results of SSB-less for inter-band CA}
  \label{fig_6}
  \end{center}
\end{figure}

Simulation results of network energy savings gains and average UE power consumption for all different resource utilization (RU) are shown Fig.\ref{fig_6}. 
As shown in Fig.\ref{fig_6}, the SSB-less SCell scheme can obtain 4.3\%$\sim $22.6\% energy saving gain in the case RU = 4.9\%$\sim $37.5\%. It can be also observed that the SSB-less SCell can provide higher UPT due to the SSB overhead reduction.

\section{Conclusion}\label{sec5}
In this article, we provide an overview of flexible spectrum orchestration for enhanced CA being discussed in 5G-Advanced. The motivation includes support of new use cases such as UL-heavy traffic, support of no DL synchronization signal transmissions for efficient use of scattered spectrum resources with narrow bandwidth and support of UL-only Scell to satisfy the regulations etc. Solutions, requirements and potential specification impacts are analyzed, and some Release 18 techniques applicable to the new CA framework are also discussed. The analysis presented in this article will hopefully shed light on better utilization of 5G spectrum and open venues for new technical proposals.

\newpage
\vspace{11pt}

\begin{IEEEbiographynophoto}{Xianghui Han} received his M.S. degree in Communication and Information System in Beijing University of Posts and Telecommunications in 2015. Since then he has been working in ZTE Corporation as a standard engineer. Meanwhile, he is currently working toward his Eng.D. in Electronic Information from Southeast University. He actively participates in standardization work in 3GPP RAN1 Working Group, involving in 5G topics such as ultra-reliable and low latency communications, coverage enhancement, sub-band full duplex operation and dynamic spectrum sharing. He is the RAN1 Feature Lead of several enhancements in Rel-17 and Rel-18. 
\end{IEEEbiographynophoto}

\begin{IEEEbiographynophoto}{Chunli Liang} received M.S. degree from Xiamen University (XMU), China, in 2005. She has been working in ZTE Corporation as a pre-research engineer since 2006. Her research interests include synchronization signal, physical uplink channel design, carrier aggregation, full duplex operation, enhancement of ultra-reliable and low latency, coverage enhancements and multi-carrier operation.
\end{IEEEbiographynophoto}

\begin{IEEEbiographynophoto}{Ruiqi (Richie) Liu} is currently a master researcher in ZTE Corporation. He serves as the Vice Chair of ISG on RIS in ETSI, the co-rapporteur of the work item on NR RRM enhancement in 3GPP. He is the Standards Liaison Officer for IEEE ComSoc Signal Processing and Computing for Communications Technical Committee (SPCC-TC) where he received the Outstanding Service Award in 2022.
\end{IEEEbiographynophoto}

\begin{IEEEbiographynophoto}{Xingguang Wei} received the M.S. degree from the Beijing University of Posts and Telecommunications (BUPT), China, in 2018. Since then he has been working in ZTE Corporation as a standardization engineer. He has been involved in the 3GPP 5G standardization since the first NR release, i.e., Rel-15, mainly focusing on bandwith part, carrier aggregation, multicast/broadcast and full duplex.
\end{IEEEbiographynophoto}

\begin{IEEEbiographynophoto}{Mengzhu Chen} received the bachelor’s degree in optical information science and technology from Wuhan University, Wuhan, China, in 2013, and the master’s degree in optical engineering from Tsinghua University, Beijing, China, in 2016. She is currently a Senior Engineer with the Department of Algorithms, ZTE Corporation. She is also a 3GPP RAN1 Delegate. Her current research interests are in the fields of 5G networks, including power saving, channel coding, interference management, and AR/VR. 
\end{IEEEbiographynophoto}

\begin{IEEEbiographynophoto}{Yu-Ngok Ruyue Li} is currently the Technical VP of Radio System Research and Standardization at ZTE Corporation, in charge of research and standardization on radio system related 5G/6G technologies. He received his Ph.D. and B.Eng. degrees from the University of Hong Kong and his M.S. degree from Stanford University.  Since he joined ZTE in 2009, he has been actively involved in wireless research and standardization activities including 3GPP standardization, with over 100 granted patents. Prior to ZTE, he worked for several telecommunication and semiconductor companies including Qualcomm and Marvell Semiconductor on projects related to baseband algorithm design and LTE standardization. 
\end{IEEEbiographynophoto}

\begin{IEEEbiographynophoto}{Shi Jin} is currently with the faculty of the National Mobile Communications Research Laboratory, Southeast University. His research interests include space time wireless communications, random matrix theory, and information theory. He serves as an Associate Editor for the  IEEE Transactions on Communications, IEEE Transactions on Wireless Communications, and IEEE Communications Letters, and IET Communications.

\end{IEEEbiographynophoto}


\begin{thebibliography}{1}
\bibliographystyle{IEEEtran}

\bibitem{ref1}
H. Ji, S. Park, J. Yeo, Y. Kim, J. Lee and B. Shim, "Ultra-Reliable and Low-Latency Communications in 5G Downlink: Physical Layer Aspects," \textit{IEEE Wireless Communications}, vol. 25, no. 3, pp. 124-130, JUNE 2018.

\bibitem{ref2}
W. Saad, M. Bennis and M. Chen, "A Vision of 6G Wireless Systems: Applications, Trends, Technologies, and Open Research Problems," \textit{IEEE Network}, vol. 34, no. 3, pp. 134-142, May/June 2020.

\bibitem{ref3}
A. Ghosh, R. Ratasuk, B. Mondal, N. Mangalvedhe and T. Thomas, "LTE-advanced: next-generation wireless broadband technology [Invited Paper]," \textit{IEEE Wireless Communications}, vol. 17, no. 3, pp. 10-22, June 2010.

\bibitem{ref4}
Nidhi, B. Khan, A. Mihovska, R. Prasad and F. J. Velez, "A Study on Cross-Carrier Scheduler for Carrier Aggregation in Beyond 5G Networks," in \textit{2022 3rd URSI Atlantic and Asia Pacific Radio Science Meeting (AT-AP-RASC)}, Gran Canaria, Spain, 2022, pp. 1-4.

\bibitem{ref5}
“TS38.213, Physical layer procedures for control.” [Online].
Available: https://portal.3gpp.org/desktopmodules/Specifications/
SpecificationDetails.aspx?specificationId=3215

\bibitem{ref6}
A. Salah, C. -H. Li, M. Al-Imari, J. Nemeth and W. Wu, "Dynamic Cross-Carrier Enhancement for 5G and B5G," in \textit{2021 IEEE 93rd Vehicular Technology Conference (VTC2021-Spring)}, Helsinki, Finland, 2021, pp. 1-6.

\bibitem{ref7}
F. Foukalas, R. Shakeri and T. Khattab, "Distributed Power Allocation for Multi-Flow Carrier Aggregation in Heterogeneous Cognitive Cellular Networks," \textit{IEEE Transactions on Wireless Communications}, vol. 17, no. 4, pp. 2486-2498, April 2018.

\bibitem{ref8}
“RP-211664, Moderator’s summary for discussion [RAN93e-
R18Prep-14] Additional RAN1/2/3 candidate topics, Set 1.”
[Online]. Available: https://portal.3gpp.org/ngppapp/CreateTdoc.aspx?
mode=view\&contributionId=1251966

\bibitem{ref9}
U. K. Dutta, M. A. Razzaque, M. Abdullah Al-Wadud, M. S. Islam, M. S. Hossain and B. B. Gupta, "Self-Adaptive Scheduling of Base Transceiver Stations in Green 5G Networks," \textit{IEEE Access}, vol. 6, pp. 7958-7969, 2018.

\bibitem{ref10}
K. -C. Chang, K. -C. Chu, H. -C. Wang, Y. -C. Lin and J. -S. Pan, "Energy Saving Technology of 5G Base Station Based on Internet of Things Collaborative Control," \textit{IEEE Access}, vol. 8, pp. 32935-32946, 2020.

\bibitem{ref11}
“RP-213554, New SI: Study on network energy savings for NR.”
[Online]. Available: https://portal.3gpp.org/ngppapp/CreateTdoc.aspx?
mode=view\&contributionId=1284708

\bibitem{ref12}
“TS38.101-1, User Equipment (UE) radio transmission
and reception; Part 1: Range 1 standalone.” [Online].
Available: https://portal.3gpp.org/desktopmodules/Specifications/
SpecificationDetails.aspx?specificationId=3283

\bibitem{ref13}
“TS38.133, Requirements for support of radio resource management.”
[Online]. Available: https://portal.3gpp.org/desktopmodules/
Specifications/SpecificationDetails.aspx?specificationId=3204

\bibitem{ref14}
“RP-220834, Revised WID on Multi-carrier enhancements.”
[Online]. Available: https://portal.3gpp.org/ngppapp/CreateTdoc.aspx?
mode=view\&contributionId=1316874

\end{thebibliography}
\end{document}